\documentclass[12pt,preprint]{aastex}

\begin{document}
\title
{Testing gravity using the growth of large scale structure in the Universe
}

\author
{E. Jennings\altaffilmark{1,2}, C. M. Baugh\altaffilmark{1},  S. Pascoli\altaffilmark{2}}
\email{elise.jennings@durham.ac.uk}
\affil{ Institute for Computational Cosmology, Department of Physics,  Durham  University , South Road, Durham, DH1 3LE, U.K.}
\affil{ Institute for Particle Physics Phenomenology, Department of Physics, Durham  University , South Road, Durham, DH1 3LE, U.K.}




\begin{abstract}
Future galaxy surveys hope to distinguish between
the dark energy and modified gravity scenarios
for the accelerating expansion of the Universe using
the distortion of clustering in redshift space. The
aim is to model the form and size of the distortion
to infer the rate at which large scale structure grows. 
We test this hypothesis and assess the performance
of current theoretical models for the redshift space
distortion using large volume N-body simulations
of the gravitational instability process. We simulate
competing cosmological models which have identical expansion
histories - one is a quintessence dark energy model with a
scalar field and the other is a modified gravity model
with a time varying gravitational constant - and demonstrate
that they do indeed produce different redshift space distortions.
This is the first time this approach has been verified using
a technique that can follow the growth of structure at the
required level of accuracy. Our comparisons show that
theoretical models for the redshift space distortion based
on linear perturbation theory give a surprisingly poor 
description of the simulation results. Furthermore, the 
application of such models can give rise to catastrophic 
systematic errors leading to incorrect interpretation 
of the observations. We show that an improved model is able 
to extract the correct growth rate. Further enhancements 
to theoretical models of redshift space distortions,
calibrated against simulations, are needed to fully
exploit the forthcoming high precision clustering measurements.
\end{abstract}

\keywords{
Methods: numerical --- Cosmology: theory --- dark energy}

\section{Introduction}
\noindent
The accelerating expansion of the Universe can be explained
either by a dark energy component or a modification to gravity.
In both alternatives, the cosmic expansion history
can be described using an effective equation of state, $w(a)$,
where $a$ is the scale factor. If two models have the same
$w(a)$, then, as a consequence, it is not possible to
distinguish them using a measurement of the expansion
history alone. Structures are, however, expected to 
collapse under gravity at different rates in dark energy 
and modified gravity cosmologies.
In general relativity, the growth of density perturbations 
depends only on the expansion history through the Hubble 
parameter, $H(a)$, or equivalently, $w(a)$
\citep{Linder:2005in}. This is not the case in modified 
gravity theories. By using the measured expansion history to 
predict the growth rate of structure and comparing this estimate 
to a direct measurement, it has been argued that it is possible 
to determine the physical origin of the accelerating cosmic expansion 
\citep{PhysRevD.69.124015,Linder:2005in}.
If there is no discrepancy between the observed growth rate and 
the prediction assuming general relativity, this implies 
that a dark energy component is responsible for the accelerated expansion.

Here we test this hypothesis using large N-body simulations
which are the only way to accurately follow the growth of cosmic 
structure and hence to probe the limits of perturbation theory.
Previous simulations of gravitational instability in 
hierarchical cosmologies have shown that linear theory gives a 
surprisingly poor description of fluctuation growth and the 
redshift space distortion of clustering, even on large 
scales \citep[e.g.][]{Angulo:2007fw, 
2008PhRvD..78b3523S,2010arXiv1003.4282J}. We simulate the growth of structure 
in a modified gravity model and a dark energy model which,  
by construction, have the same expansion history.
The growth rate is measured from the appearance of the power 
spectrum in redshift space. The goals of this paper 
are, firstly, to determine if these competing 
cosmologies can be distinguished from the distortion of clustering 
as measured in redshift space, using the simulation results, and 
secondly, to test theoretical models of the power 
spectrum in redshift space against the simulation results, to assess 
how well they can recover the growth rate.

This letter is set out as follows. In Section \ref{thecos} we 
review the growth of perturbations and describe 
the modified gravity model. Clustering in redshift space is 
measured in Section 3, and theoretical models are applied 
to describe the simulation results. In Section 4 we present our conclusions.

\section{The cosmological models and simulations \label{thecos}}

Here we recap how perturbation growth depends on the expansion history 
and the strength of gravity (Section \ref{TestGR}), before outlining 
the modified gravity model (Section \ref{1.1}) and our N-body simulations 
(Section \ref{mgsim}).

\subsection{The linear growth rate \label{TestGR}}

In the framework of general relativity (GR), the growth of a density
fluctuation, $\delta \equiv (\rho(x,t)-\bar{\rho}_{\rm m})/\bar{\rho}_{\rm m}$, where $\bar{\rho}_{ \rm m}$ is the average matter density, 
depends only on the expansion history, $H(a)$.
Using the perturbed equations of motion, within GR, 
the growth of perturbations follows
\begin{eqnarray}
\label{delta}
\ddot{\delta} +2H\dot{\delta} -4\pi G_N \rho_{\rm m} \delta = 0 \, ,
\end{eqnarray}
where $G_N$ is the present gravitational constant found in laboratory 
experiments and a dot denotes a time derivative.
The growth rate is $f \equiv {\rm d ln}\delta/{\rm d ln} a$, where
$\delta(a)$ is the growing mode solution to Eq. \ref{delta}.
Changing variables to $g \equiv \delta/a$ and allowing the 
gravitational constant to vary in time, denoted by $\tilde{G}$, 
gives \citep{Linder:2005in}
\begin{eqnarray}
\label{linder}
\frac{{\rm{d}}^2g}{{\rm{d}}a^2}  &+& \left( 5 + \frac{1}{2} \frac{{\rm{d ln}}H^2}{{\rm{d ln}} a}\right)\frac{1}{a}\frac{{\rm{d}}g}{{\rm{d}}a}  \\ \nonumber
&+& \left( 3 + \frac{1}{2}\frac{{\rm{d ln}}H^2}{{\rm{d ln}} a} -\frac{3}{2}\frac{\tilde{G}(a)}{G_N}\Omega_{\rm{m}}(a) \right)g = 0 \, ,
\end{eqnarray}
where $\Omega_{\rm{m}}(a)$ is the matter density parameter. 
Eq. \ref{linder} shows that in GR, 
$\tilde{G}(a)/G_N =1$ and the growth of perturbations depends only 
on the expansion history, $H(a)$. In modified gravity theories, 
however, the growth of perturbations depends on both $H(a)$ 
and $\tilde{G}(a)$.

\subsection{Time variation of Newton's constant \label{1.1}}

Modifications to GR provide an alternative 
explanation to dark energy for the accelerating cosmic expansion.
Modified gravity theories can generally be divided into models 
which introduce a new scalar degree of freedom to Einstein's 
equations, e.g. scalar tensor or $f(R)$ theories, and those 
which change dimensionality of space, e.g. braneworld gravity.
In many such models, the time variation of fundamental 
constants, such as Newton's gravitational constant, $G_N$, 
is naturally present. 

Self consistent scalar tensor theories are viable alternatives 
to GR and give rise to an accelerating expansion 
at late epochs. We  refer to these as \lq extended quintessence\rq \, models.
Calculations which follow spatial variations in the scalar field 
have shown that, in practice, a broad range of these models can 
be effectively described with a time varying Newton's constant 
\citep{2008PhRvD..77j3003P, 2010arXiv1009.1400L}.

The variation of $G_N$ is constrained by various observations, such 
as the lifespan of stars \citep{PhysRev.73.801}, the age of globular 
clusters \citep{1996A&A...312..345D}, the mass of neutron stars 
\citep{1996PhRvL..77.1432T} and the synthesis of light nuclei
\citep{2005PhRvD..72d4010U, 2005PhRvD..71l3526C}.
A time-varying $G_N$ would also modify the temperature fluctuations 
in the cosmic microwave background, shifting the peaks  to 
larger (smaller) scales on increasing (decreasing) $G_N$. 
This leads to a constraint on the variation of $G$, 
$\dot{G}/G  = (-9.6 \sim 8.1) \times 10^{-12}$ yr$^{-1}$ 
\citep{2007PhRvD..75h3521C}.

\begin{figure*}
\center
\epsscale{1.}
\plotone{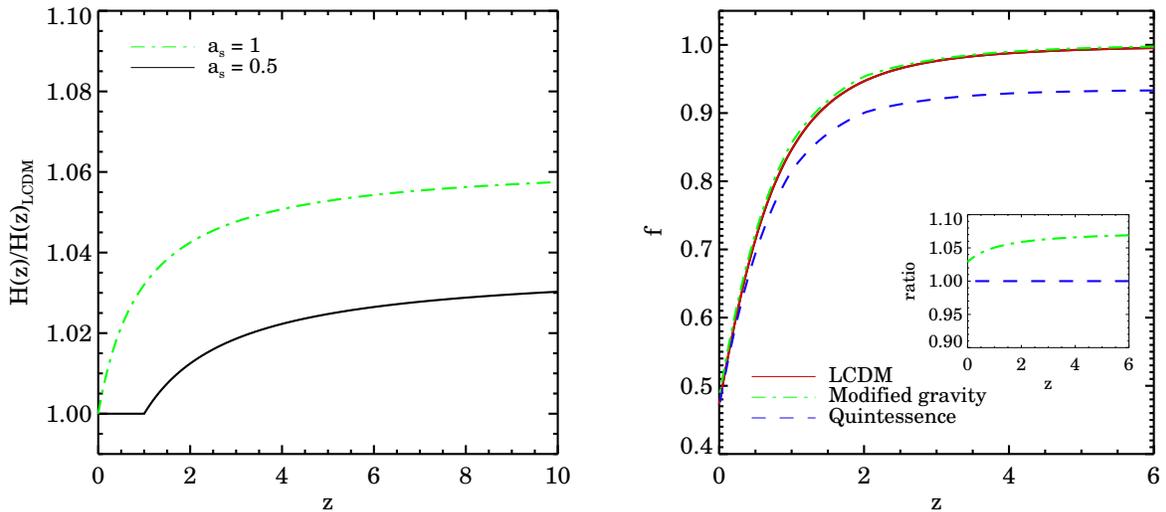}
\caption
{Left: Ratio of the expansion rate, $H(z)$, to 
that in $\Lambda$CDM for two modified gravity models 
specified by Eq.~\ref{GG}: dot dashed green line: $a_s = 1,
\mu^2_0 = 1.13$ and solid black line : $a_s = 0.5, \mu^2_0 = 1.075$.
Right: The linear growth rate, $f$, as a function of redshift 
for $\Lambda$CDM (red solid), a modified gravity cosmology
with $a_s=1$ and $\mu^2_0=1.13$ (green dot dashed) and a 
quintessence model (blue dashed) with the same expansion history 
as the modified gravity model.
The inset shows the ratio of $f$ in the modified gravity model
to that in the quintessence model as a function of redshift
(green dot dashed line).
}\label{mgFig1}
\end{figure*}

Here we consider a simple model 
for $\tilde{G}$ 
\citep{2003PhRvD..67f3002Z,2005PhRvD..72d4010U, 2007PhRvD..75h3521C},
\begin{eqnarray}
\label{g2}
\tilde{G} &=& \mu^2 G_N \, ,
\end{eqnarray}
where
\begin{eqnarray}
\label{GG}
\mu^2 =
\left\{
\begin{array}{cr}
\mu_0^2  & \mbox{if} \, \, a<a_* \\ 
        1 - \frac{a_s-a}{a_s-a_*}(1-\mu_0^2 )  &\mbox{if} \, \, a_*\le a \le a_s \\ 
        1   & \mbox{if} \, \, a>a_s \,.
\end{array}
\right.
\end{eqnarray}
This parametrization describes a smoothly varying $\tilde{G}$ 
which converges slowly to its present value, $G_N$,
and is more physical than those based on step functions
\citep[e.g.][]{2010PhRvD..81j3528C}.
The parameter, $a_*$, denotes the scale-factor of photon decoupling 
and the parameters $\mu_0$ and $a_s$ quantify the deviation 
of $\tilde{G}$ from the laboratory measured value, $G_N$, and 
the scale factor at which $\tilde{G}$ and $G_N$ are equal, respectively.
The background evolution is given by
\begin{eqnarray}
\label{HH}
H^2 = H^2_0 \frac{\tilde{G}}{G_N} \left(\frac{ \Omega_{\rm m}}{a^3}  + \Omega_{\tiny \mbox{DE}} e^{3\int_a^1 {\rm{d ln}} a' [1+w(a')]}\right) \, .
\end{eqnarray}
Note we assume an equation of state
$w(a)= -1$ in the modified gravity model 
to match $\Lambda$CDM. 
In Eq. \ref{HH},  $\Omega_{\tiny \mbox{DE}} $ is the 
ratio of the dark energy density to the critical density today. 
In the left panel of  Fig. \ref{mgFig1}, 
we plot the ratio of the Hubble rate for two different cosmological models with varying $\tilde{G}$, to the Hubble rate in a  $\Lambda$CDM cosmology as
a function of redshift.
We chose to simulate the model with the maximum deviation of $\tilde{G}$ 
from $G_N$ which is still compatible with CMB measurements
and solar system constraints ($\tilde{G} \to G$ as $a \to 1$), which
occurs for a stabilization redshift corresponding to $a_s = 1$ 
(i.e. the green dot dashed line in Fig.~\ref{mgFig1}).

\subsection{N-body simulations \label{mgsim}}

We use large volume N-body simulations to carry out 
the first direct test of the idea that dark energy 
and modified gravity cosmologies which, by construction, have exactly
the same expansion history, can be distinguished by 
a measurement of the rate at which structure grows. 
The modified gravity model we simulate 
has the maximum deviation from Newton's 
constant that is compatible with observational constraints, 
as discussed above.
We construct a quintessence model by fitting the 
 expansion history to match the varying $\tilde{G}$ model 
within 0.25\% over $0 \le z \le 200$.
This model is consistent with constraints on dynamical dark 
energy \citep{Komatsu:2008hk,2009MNRAS.400.1643S}.

The simulations were carried out using a memory efficient version of 
the  TreePM code {\tt Gadget-2}, called {\tt L-Gadget-2} 
\citep{Springel:2005mi}.
The simulation used $N=1024^3 \sim 1 \times 10^9$ particles  
in a box of comoving length $1500 h^{-1}$Mpc. The comoving 
softening length was $\epsilon = 50 h^{-1}$kpc and the present 
day linear rms fluctuation in spheres of radius 
8 $h^{-1}$ Mpc is  $\sigma_8 = 0.8$.
Simulations of extended quintessence cosmologies 
need to account for both the gravitational correction 
due to a varying $\tilde{G}$ in the Poisson equation and a modified 
expansion history \citep[see][]{2008PhRvD..77j3003P}.
In the modified gravity simulation, both the long and short-range  
TreePM algorithm force computations are modified to include a 
time-dependent gravitational constant.
In both the dark energy and modified gravity simulations the Hubble 
parameter computed by the code was also changed 
as in \citet{2010MNRAS.401.2181J}.

The linear theory power spectrum used to generate the initial
conditions was  obtained using CAMB \citep{Lewis:2002ah}.
We adopt a $\Lambda$CDM linear theory power spectrum at $z=0$, 
and use consistent linear growth factors in each cosmology to 
obtain the power spectrum amplitude at $z=200$. 
In principle, as the quintessence cosmology could be classed as an 
early dark energy model, the linear theory spectrum should be 
modified in shape. However, as we have shown, such a change has 
a negligible impact on the nonlinear spectrum and on the 
ratio of the quarupole to monopole monents (Jennings et~al. 2010a,b).

To obtain errors on our measurements we  ran 10 lower resolution 
simulations with $512^3$ particles, also in a box of
comoving length $1500 h^{-1}$Mpc, with different realizations 
of the density field.
The power spectrum was computed using the cloud in cell (CIC) 
assignment scheme and performing a fast Fourier transform.
For the initial conditions the linear growth rate for each model and $\Lambda$CDM was obtained by solving Eq.~\ref{linder} numerically
and is plotted in the right hand panel of Fig.~\ref{mgFig1} 
as a function of redshift.
For  all the models we used the following cosmological parameters:
$\Omega_{\rm m} = 0.26$,
$\Omega_{\rm{DE}}=0.74$, $\Omega_{\rm b} = 0.044$,
$h_0 = H_0/100$km s$^{-1}$Mpc$^{-1} = 0.715$ and a spectral index of $n_{\mbox{s}} =0.96$ \citep{2009MNRAS.400.1643S}.
We have verified that  our modifications to Gadget-2  are accurate 
by checking that the growth of the fundamental mode in the simulations 
agrees with the linear theory predictions.
\begin{figure*}
\center
\epsscale{1.5}
\plotone{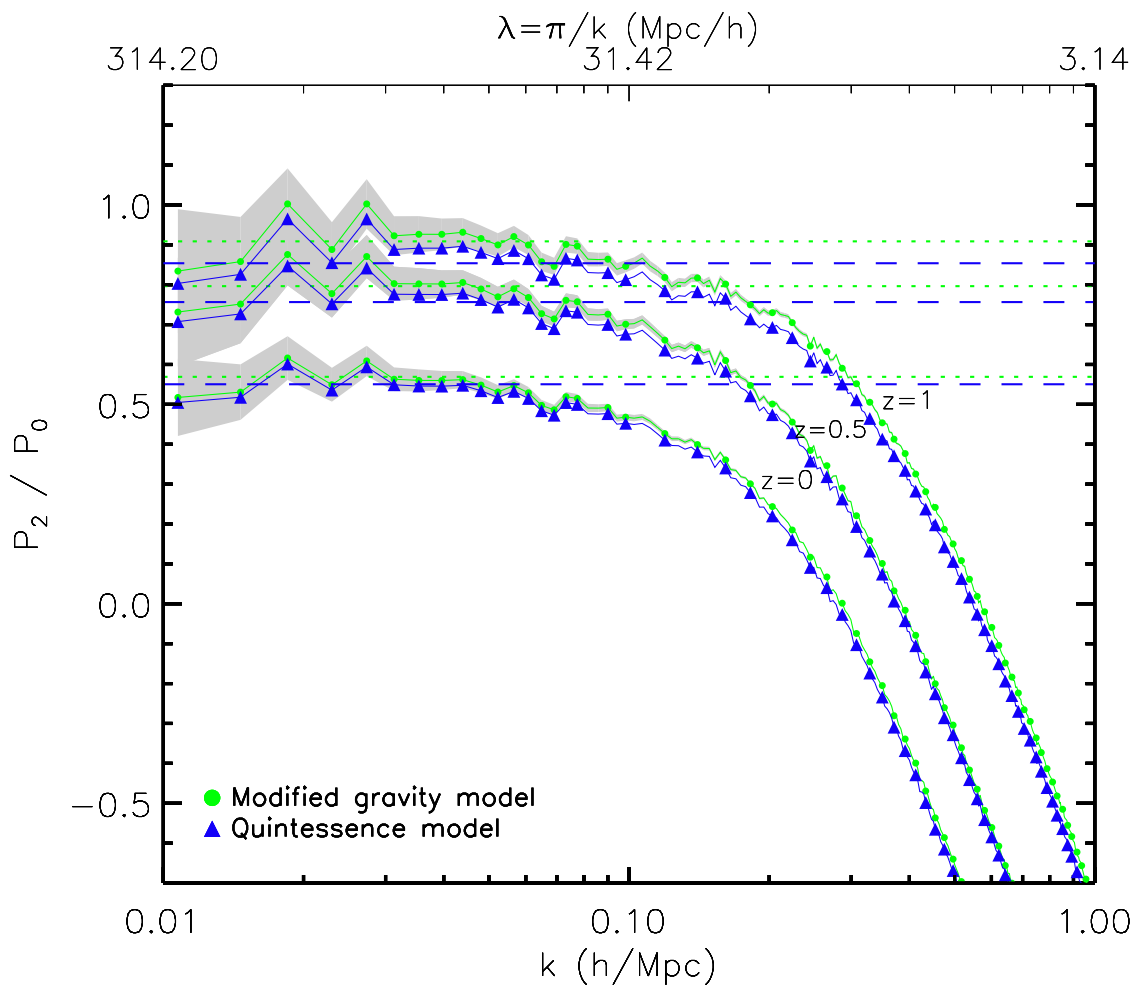}
\caption
{
The ratio of the quadrupole and monopole moments of the redshift 
space power spectrum, $P_{2}/P_{0}$, as a function of wavenumber 
at redshifts  $z=0$, 0.5 and 1 (in order of ascending amplitude).
The points show the N-body results, for the modified gravity model 
(green triangles) and the quintessence model (blue circles). 
The shading indicates the error on the ratio, estimated from the 
scatter over 10 lower resolution simulations. The horizontal lines
show the predictions of linear theory model, with the colours 
having the same meaning as those used for the points.
}\label{mgFig3}
\end{figure*}

\section{Results \label{MGresults}}

We now briefly recap the models used to describe 
the redshift space distortion of the matter power spectrum and 
then (S~\ref{mgf})fit these models to the moments of the 
power spectrum measured in our simulations.

\subsection{Redshift space distortions \label{mgrsd}}
\begin{figure*}
\center
\epsscale{1.2}
\plotone{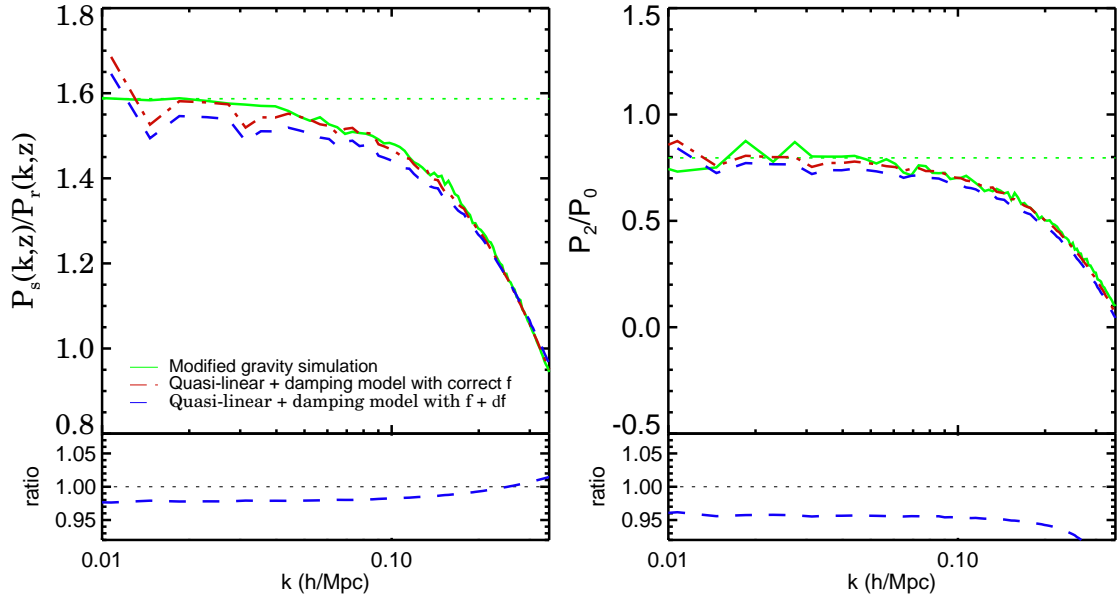}\vspace{-5cm}
\caption{
Left: The ratio of the monopole of the redshift 
space power spectrum to the real space $P(k)$ at $z=0.5$, 
as a function of wavenumber.
Right: The ratio of the quadrupole to monopole moment of 
the redshift space $P(k)$.
The quasi-linear plus damping model is plotted using 
$f=f_{\tiny \mbox{true}}$ ($f=1.05 f_{\tiny \mbox{true}}$) 
as a red dot dashed (blue dashed) line.
The lower panels show the ratio of $P^s_0/P_r$ (right: $P^s_2/P^s_0$) 
using $f= 1.05 f_{\tiny \mbox{true}}$ to the same model 
using $f=f_{\tiny \mbox{true}}$ (blue dashed line). 
}\label{fig3new}
\end{figure*}
\begin{figure*}
\center
\epsscale{1.2}
\plotone{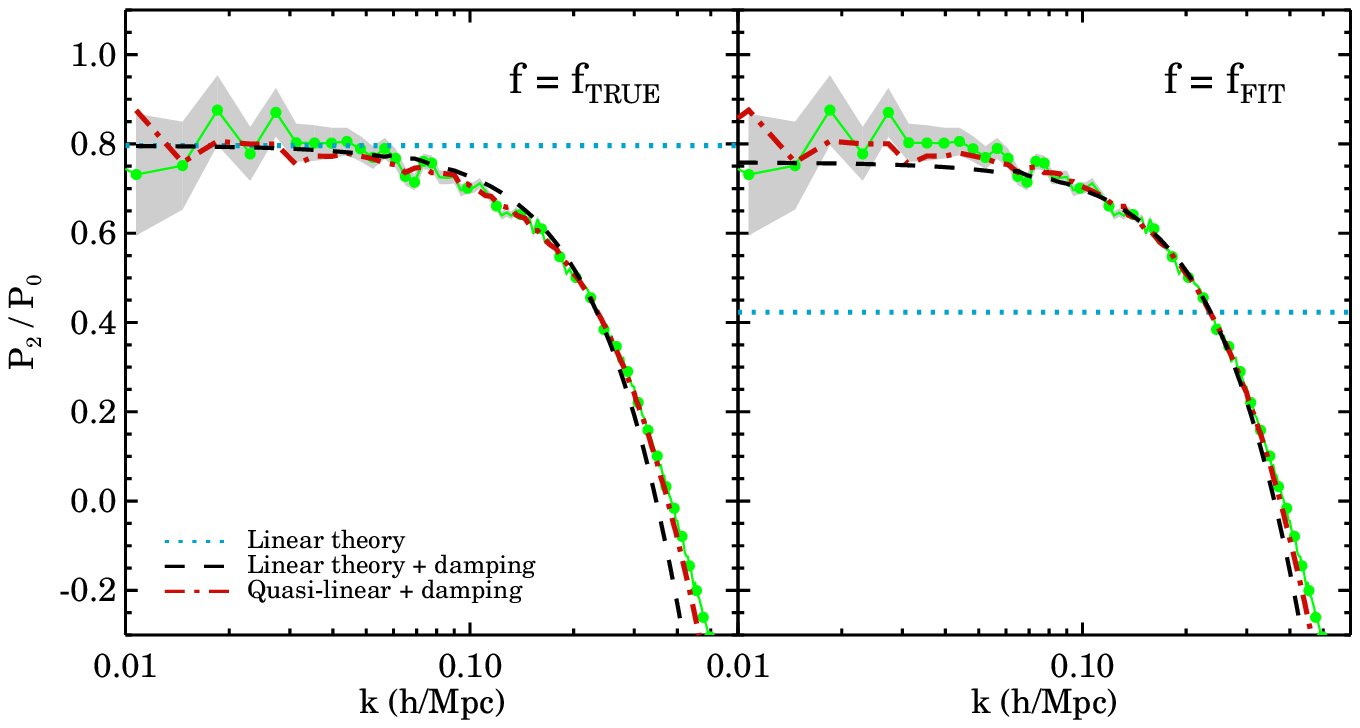}
\caption
{
The ratio  $P_2/P_0$ in the modified gravity cosmology
measured from the high resolution simulation (green points),
together with three models for $P_2/P_0$, using the correct linear
growth rate, $f=f_{\tiny \mbox{TRUE}}$ (left), and the value
of $f$ obtained in a $\chi^2$ fit over the wavenumber range
$0.01 \le k (h/$Mpc$) \le 0.25$,
$f=f_{\tiny \mbox{FIT}}$ (right).
The shaded region shows the propagated errors from ten
lower resolution simulations. The models plotted are indicated
by the key: linear theory - blue dotted line, linear theory plus
damping - black dashed line and quasi-linear plus damping - red dashed line.
In the left panel the best fit value for $\sigma_p$ ($\sigma_v$)
obtained in the range $0.01 \le k (h/$Mpc$) \le 0.25$,with fixed $f$,
was used for the  linear theory plus damping
(quasi-linear plus damping) model.
}\label{3.3}
\end{figure*}

The matter power spectrum in redshift space can be decomposed 
into multipole moments using Legendre polynomials.
The ratio of the quadrupole and monopole moments of the 
matter power spectrum is plotted in Fig.~\ref{mgFig3}. 
We model redshift space distortions in the distant observer
approximation by perturbing the particle positions down
one of the cartesian axes, using the suitably scaled
component of the peculiar velocity.
The simulation results show that this ratio has a strong 
dependence on wavenumber. This can be contrasted with  
the linear perturbation theory prediction 
\citep{Cole:1993kh}, 
\begin{eqnarray} 
\frac{P_2(k)}{P_0(k)} &=&  \frac{4\beta/3 +4\beta^2/7}{1 + 2\beta/3 + \beta^2/5}\, , 
\label{lint}
\end{eqnarray}
where $\beta=f/b$ and $b$ is the linear bias, which is unity 
for dark matter; Eq.~\ref{lint} is independent of scale
(horizontal lines in Fig.~\ref{mgFig3}). 
We note that, by considering redshift space distortions in 
the clustering of the dark matter, we are testing theoretical 
models against the simplest possible case. The distortions will 
inevitably be more complicated for dark matter haloes and galaxies, 
for which the bias factor $b$ can have scale dependence 
(e.g. Angulo et~al. 2008). 

In Fig ~\ref{mgFig3} the quadrupole to 
monopole ratio increases in amplitude with redshift, due to the
evolution in the matter density parameter. At $z=0$ there 
is a  ~2.5\% difference between the linear theory growth rates 
in the two models. However, at this level, the measured ratios $P_{2}/P_{0}$
are indistinguishable on the very largest scales $k<0.02 h/$Mpc 
(green dotted and blue dashed horizontal lines). 
At $z=0.5$ and $z=1$ the linear theory predictions for the growth rates 
in the two models differ by 4\% and 6\% respectively.
The error on this ratio measured from the lower resolution simulations 
is shown by the shaded region in Fig. \ref{mgFig3}.

\begin{figure*}
\center
\epsscale{1.1}
\plotone{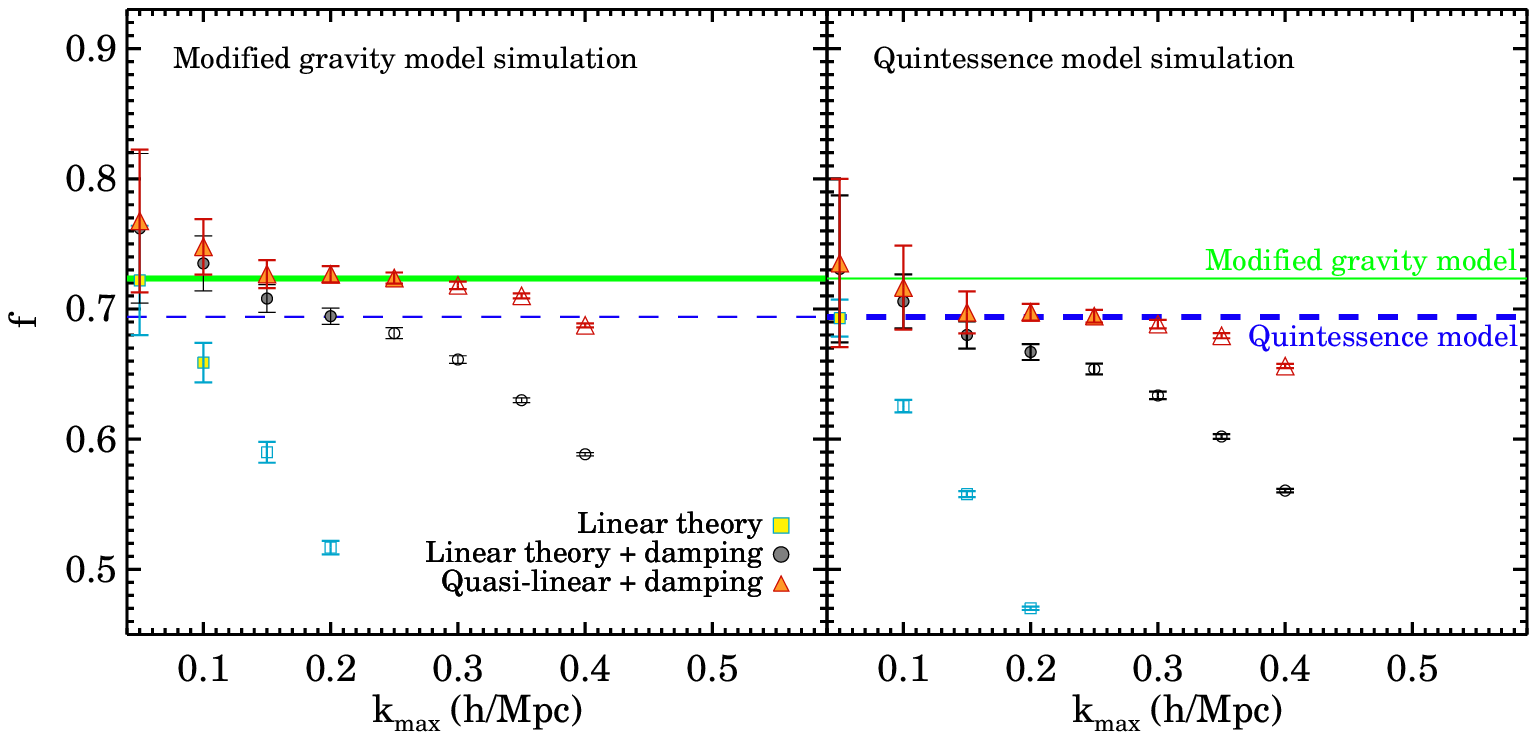}
\caption
{
Measurements of the growth rate $f$.
The results are plotted as a function of the maximum wavenumber 
used in the fit, $k_{\tiny \mbox{max}} (h/$Mpc). The symbols
show the results of fitting to $P_{2}/P_{0}$ at $z=0.5$
using different models: linear theory - squares, linear theory 
plus damping - circles, quasi-linear plus damping - triangles.
The symbols are filled on scales where the model is a good
description of the measured ratio. The error bars represent
the 1$\sigma$ uncertainty. Left: we fit to the
modified gravity model and aim to recover the true growth factor
shown by the thick green horizontal line. Right: we fit to the 
quintessence model, with the target growth factor
shown by the thick blue dashed line.
}\label{mainFig2}
\endcenter
\end{figure*}

In addition to the linear theory model we consider two variants. 
The first is the Gaussian model \citep{1994MNRAS.267.1020P},
\begin{eqnarray}
P^s(k,\mu) =  P^r(k) (1+\beta \mu^2)^2 e^{(-k^2 \mu^2 \sigma_p^2)} \, ,
\label{gaussianmodel}
\end{eqnarray}
where $\sigma_p$ is the pairwise velocity dispersion along the line of 
sight, which is treated as a parameter to be fitted.
We refer to Eq. \ref{gaussianmodel} as the ``linear theory plus damping'' 
model.
The damping introduces a scale dependence into 
the ratio $P_{2}/P_{0}$. The second variant model takes into
account departures from linear theory, as well as including 
small scale damping
\citep{Scoccimarro:2004tg}: 
\begin{eqnarray}
\label{SM}
P^s(k,\mu)= &&( P_{\delta \delta}(k) + 2 f\mu^2 P_{\delta \theta}(k)               \\ 
&& + f^2\mu^4P_{\theta \theta}(k)) 
\times e^{-(f k \mu \sigma_v )^2} \, , \nonumber
\end{eqnarray}
where   $\sigma_v$ is the 1D linear velocity dispersion 
and $P_{\theta \theta}$ and $P_{\delta \theta}$ are the velocity 
divergence auto and cross power spectrum respectively measured from the simulations \citep[see also][]{2010arXiv1003.4282J}.
We refer to Eq.~\ref{SM} as the  ``quasi-linear plus damping'' model.
 We note that $P_2/P_0$ is more sensitive to changes in $f$ than
the ratio of the monopole moment of the redshift space to real
space $P(k)$, see Fig.~\ref{fig3new},  and, as a result, 
the 1 $\sigma$ errors for $f$ are smaller when fitting 
to $P_2/P_0$.

\subsection{Measuring the growth rate \label{mgf}}

We now apply the above models to the simulation results. 
In Fig. \ref{3.3}, we plot the measured ratio $P_2/P_0$, for 
the modified gravity simulation at $z=0.5$, together 
with the theoretical predictions.
In the left panel, the correct value of $f$ for this 
cosmology together with the best fit value for $\sigma_p$ 
and $\sigma_v$ in the range $0.01 \le k (h/$Mpc$) \le 0.25$ was
used in the linear theory plus damping and quasi-linear plus 
damping models respectively.
In the right panel, the best fit value for $f$ obtained by 
fitting over the same range of wavenumbers is used for all 
models plotted.
The value of $f$ obtained for the linear theory model 
is sensitive to the maximum value of $k$ used in the fit. 
It is clear that both the linear theory and the linear theory plus 
damping models fail to predict the correct value for $f$, with 
the best fitting values differing by $\sim40$\% and
$\sim 6$\% respectively from the true value,
 see Fig. \ref{mainFig2}. 
All the models plotted in the right panel in Fig.~\ref{3.3} 
use the value of $f$ recovered when $k_{\tiny \mbox{max}} = 0.25h/$Mpc. 
The quasi-linear plus damping model recovers the correct 
value of $f$ over this wavenumber range to within $\sim 0.64$\%.

To test these models for the redshift space power spectrum further 
we vary the maximum wavenumber, $k_{\rm max}$, used in the fit and 
plot the recovered growth rate as a function of $k_{\rm max}$ in 
Fig.~\ref{mainFig2}. With  an accurate model we would
recover the correct value for the growth rate $f$ and the 
answer would be independent of the value of $k_{\rm max}$ adopted, 
with the only sensitivity to $k_{\rm max}$ being in the error on 
the growth rate. Fig.~\ref{mainFig2} shows
that the quasi-linear plus damping model comes closest 
to meeting this ideal. Even this model breaks down beyond 
$k_{\rm max} \sim 0.3 h /{\rm Mpc}$, which suggests that the 
modelling of the small scale velocity dispersion can be improved. 
Most importantly, this model recovers the correct value for
$f$ and can distinguish between the two cosmologies. 
The models based on linear theory perform less well.
In fact, the answer depends strongly on the maximum wavenumber 
used in the fit. 
In Fig.~\ref{mainFig2} filled symbols are plotted for
scales over which the model is a good description of the measured ratio (i.e. 
$\chi^2/\nu \sim 1$, where $\nu$ is the number of degrees of freedom).

\section{Conclusions \label{conc}}

Forthcoming galaxy redshift surveys aim to resolve 
fundamental questions in cosmology, such as the origin 
of the accelerating expansion. 
We have measured redshift space distortions in two
simulations with different cosmologies and demonstrated 
that a modified gravity model, described by a time varying Newton's 
constant, and a dark energy model, which have identical expansion
histories, have measurably different growth rates.
We have tested models for redshift space distortions including 
commonly used linear theory models. 
We find that models based on linear theory fail to recover 
the correct value of the growth rate. A quasi-linear model 
including non-linear velocity divergence terms is far more accurate and 
allows us to distinguish between these competing cosmologies.

Even though we consider large scales, there are 
important departures from linear theory which can only be 
modelled by N-body simulation \citep{2010arXiv1003.4282J}. 
Without such guidance, the application of models based on 
linear theory could lead to systematic errors of the same 
order as the difference in $f$ between competing cosmologies. 
In this event, such models would give the wrong conclusion 
about the physics driving the cosmic acceleration.
We find that an improved model is able to recover the correct 
growth factor and hence to tell the models apart. This model 
can be applied to the measured power spectrum over a wider 
range of scales than those based on linear theory, making 
better use of the available data. Our tests show  
that a further improvement to this model is possible. 
Nevertheless our results show that with such improved 
models validated against simulations, the prospects of 
distinguishing between modified gravity and dark energy 
using clustering measurements are encouraging.

\section*{Acknowledgments}
EJ acknowledges a fellowship from the European
Commission's Framework Programme 6, through the Marie Curie Early Stage
Training project MEST-CT-2005-021074.
This work was supported by the Science and Technology Facilities
Council.


\label{lastpage}

\end{document}